\documentclass[english,aps, prb, twocolumn]{revtex4-1}
\usepackage[T1]{fontenc}
\usepackage[latin9]{inputenc}
\usepackage{amsmath}
\usepackage{amssymb}
\usepackage{graphicx}

\makeatletter

\@ifundefined{textcolor}{}
{%
 \definecolor{BLACK}{gray}{0}
 \definecolor{WHITE}{gray}{1}
 \definecolor{RED}{rgb}{1,0,0}
 \definecolor{GREEN}{rgb}{0,1,0}
 \definecolor{BLUE}{rgb}{0,0,1}
 \definecolor{CYAN}{cmyk}{1,0,0,0}
 \definecolor{MAGENTA}{cmyk}{0,1,0,0}
 \definecolor{YELLOW}{cmyk}{0,0,1,0}
}

\makeatother

\usepackage{babel}
\begin{document}

\preprint{This line only printed with preprint option}

\title{Measuring nonadiabaticity of molecular quantum dynamics with quantum
fidelity and with its efficient semiclassical approximation}

\author{Tomáš Zimmermann}

\author{Ji\v{r}\'{\i} Van\'{\i}\v{c}ek}

\email{jiri.vanicek@epfl.ch}

\selectlanguage{english}%

\affiliation{Laboratory of Theoretical Physical Chemistry, Institut des Sciences
et Ingénierie Chimiques, Ecole Polytechnique Fédérale de Lausanne
(EPFL), CH-1015, Lausanne, Switzerland}

\begin{abstract}
We propose to measure nonadiabaticity of molecular quantum dynamics
rigorously with the quantum fidelity between the Born-Oppenheimer
and fully nonadiabatic dynamics. It is shown that this measure of
nonadiabaticity applies in situations where other criteria, such as
the energy gap criterion or the extent of population transfer, fail.
We further propose to estimate this quantum fidelity efficiently with
a generalization of the dephasing representation to multiple surfaces.
Two variants of the multiple-surface dephasing representation (MSDR)
are introduced, in which the nuclei are propagated either with the
fewest-switches surface hopping (FSSH) or with the locally mean field
dynamics (LMFD). The LMFD can be interpreted as the Ehrenfest dynamics
of an ensemble of nuclear trajectories, and has been used previously
in the nonadiabatic semiclassical initial value representation. In
addition to propagating an ensemble of classical trajectories, the
MSDR requires evaluating nonadiabatic couplings and solving the Schr\"{o}dinger
(or more generally, the quantum Liouville-von Neumann) equation for
a single discrete degree of freedom. The MSDR can be also used to
measure the importance of other terms present in the molecular Hamiltonian,
such as diabatic couplings, spin-orbit couplings, or couplings to
external fields, and to evaluate the accuracy of quantum dynamics
with an approximate nonadiabatic Hamiltonian. The method is tested
on three model problems introduced by Tully, on a two-surface model
of dissociation of NaI, and a three-surface model including spin-orbit
interactions. An example is presented that demonstrates the importance
of often-neglected second-order nonadiabatic couplings.
\end{abstract}
\maketitle

\section{Introduction\label{sec:Introduction}}

Nonadiabatic effects give rise to a great variety of phenomena in
chemical dynamics.\cite{Butler1998,Worth2004,FDisscuss2004} To account
for these effects, many theoretical methods have been developed. The
most accurate but also the most computationally demanding are wave
packet approaches which solve the Schr\"odinger equation for both
electrons and nuclei directly. Some wave packet methods, e.g., the
multi-configuration time-dependent Hartree method,\cite{Meyer2009,Eroms2010}
have been successfully applied to problems with tens of degrees of
freedom. Trajectory based nonadiabatic Bohmian dynamics\cite{Wyatt2001,Lopreore2002}
is another in principle exact method, which can be, moreover, combined
with electronic structure computed on the fly.\cite{Curchod2011}
Less accurate but also less expensive are various semiclassical approaches,
which can also describe some quantum effects, especially on the nuclear
motion. These include multiple-spawning methods,\cite{Martinez1996,Yang2009}
methods based on the Herman-Kluk propagator,\cite{Burant2002,Kondorskiy2004,Wu2005,Miller2009}
or the surface hopping and jumping method of Heller \textit{et al.}\cite{Heller2002}
The most widely used are methods in which the nuclei are treated classically
and the quantum effects enter only through interaction with electrons,
which are described quantum mechanically. Among these belong methods
based directly on the mixed quantum-classical Liouville equation,\cite{Donoso1998,Kapral1999,Horenko2002,Ando2003,Horenko2004,Micha2004,MacKernan2008,Bonella2010,Bousquet2011}
the mean field Ehrenfest dynamics, various surface hopping methods,\cite{Tully1990,Nielsen2000,Shenvi2009}
or methods in which the classical limit is obtained by linearizing
the path integral representation of the quantum propagator.\cite{Dunkel2008}

Unfortunately, all of these methods are significantly more computationally
demanding than their adiabatic or diabatic counterparts. (In the following,
we discuss mostly nonadiabatic dynamics. Nevertheless, the discussion
holds almost entirely also for the nondiabatic dynamics in the diabatic
basis.) One goal of this paper is to find a general criterion which
would determine when a given dynamics is nonadiabatic enough to justify
the use of the expensive nonadiabatic methods. Several possible criteria
could be envisaged. One widely used criterion is the extent of population
transfer or, more precisely, the decay of survival probability $P_{\mathrm{QM}}$
on the initially populated potential energy surface (PES) due to nonadiabatic
couplings. Although a fast decay of $P_{\mathrm{QM}}$ is a clear
sign of nonadiabaticity of the dynamics, the opposite implication
is not necessarily true, as can be in seen in Ref. \onlinecite{Zimmermann2010}
and as will also be demonstrated below. Another criterion, often employed
to decide when to ``switch on'' the couplings in nonadiabatic calculations,\cite{Tao2009}
is the energy gap criterion: one simply monitors the energy difference
between PESs and when it becomes sufficiently small, the dynamics
is considered nonadiabatic. While useful in practical calculations,
this criterion does not always reflect the actual extent of nonadiabaticity,
which also depends on nonadiabatic couplings and on the nuclear momentum.
Other approximate criteria, which are intermediate between the two
criteria mentioned, estimate the change of $P_{\mathrm{QM}}$ from
basic properties of the PESs, from couplings between the PESs, and
from the nuclear velocity. Examples include variants\cite{Frauenfelder1985}
of the Landau-Zener-St\"{u}ckelberg model.\cite{Landau1932,Landau1932a,Zener1932,Stuckelberg1932,Nikitin1999}

\begin{figure*}
\includegraphics[bb=0bp 0bp 566bp 75bp,width=\linewidth]{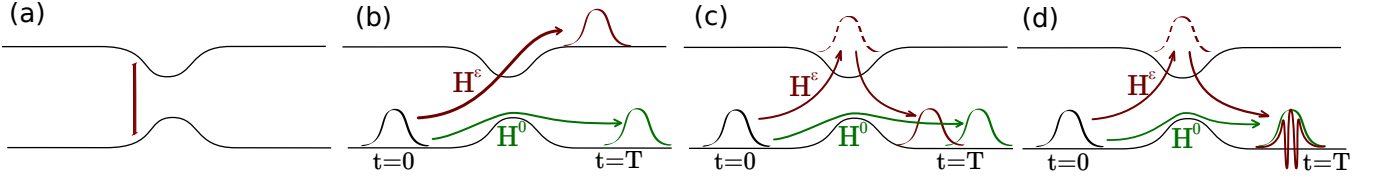}

\caption{Possible criteria of non(a)diabaticity of quantum dynamics. The static
energy-gap criterion does not take into account the dynamics of the
wave packet (a). The population transfer criterion measures the actual
decay of probability density on the initial PES (b). It is more sensitive
than the static energy-gap criterion. Fidelity criterion can capture
the population transfer (b) as well as other nonadiabatic effects
such as displacement (c) or interference (d) on a single PES, which
would be undetected by the population transfer criterion. $\hat{\mathbf{H}}^{0}$
is the decoupled Born-Oppenheimer Hamiltonian, whereas $\hat{\mathbf{H}}^{\epsilon}$
is the fully coupled nonadiabatic Hamiltonian. \label{fig:nonadiab_criteria}}
\end{figure*}

In Ref. \onlinecite{Zimmermann2010} we proposed a more rigorous quantitative
criterion of the non(a)diabaticity of the quantum dynamics, based
on the quantum fidelity $F_{\mathrm{QM}}$\cite{Peres1984} between
the adiabatically and nonadiabatically propagated molecular wave functions.
More precisely, 
\begin{equation}
F_{\mathrm{QM}}(t)=|f_{\mathrm{QM}}(t)|^{2}=\vert\langle\psi^{0}(t)\vert\psi^{\epsilon}(t)\rangle\vert^{2},\label{eq:Fidelity_definition}
\end{equation}
where $\left|\psi^{0}(t)\right\rangle $ is the quantum state of the
molecule evolved using the adiabatic Hamiltonian $\hat{H}^{0}$ with
uncoupled PESs and $\vert\psi^{\epsilon}(t)\rangle$ is the quantum
state evolved using the fully coupled nonadiabatic Hamiltonian $\hat{H}^{\epsilon}$.
When $F_{\mathrm{QM}}\approx1$, $\left|\psi^{0}(t)\right\rangle $
is close to $\left|\psi^{\epsilon}(t)\right\rangle $ and an adiabatic
simulation is a good approximation to the nonadiabatic simulation.
When $F_{\mathrm{QM}}\ll1$, adiabatic treatment is inadequate and
a nonadiabatic method should be used. Unlike the energy gap and population
transfer criteria, the fidelity criterion can detect more subtle nonadiabatic
effects caused, e.g., by the displacement and/or interference on a
single PES surface (see Fig.\,\ref{fig:nonadiab_criteria}). Panels
(c) and (d) of Fig.\,\ref{fig:nonadiab_criteria} show two extreme
examples in which the nonadiabatic couplings induce a hop to the excited
surface followed by a hop back to the ground surface so that at large
times, only the ground state is occupied. While the nonadiabatic couplings
have no effect on the survival probability on the ground surface,
they have a large effect on the molecular wavefunction since the returning
wavepacket may accumulate a time delay (panel c) or a phase (panel
d), and hence can have a zero overlap with the wavepacket propagated
adiabatically. Although neither case can be detected by the survival
probability criterion, both scenarios can be detected easily by fidelity
\eqref{eq:Fidelity_definition}.  

Due to the generality of definition \eqref{eq:Fidelity_definition},
fidelity can be employed in many applications with nonadiabatic dynamics,
depending on the choice of $\hat{H}^{0}$ and $\hat{H}^{\epsilon}$.
Very recently, this fidelity criterion of nonadiabaticity was used
to find the optimal time-dependent Hamiltonian maximizing the adiabaticity
of the dynamics from an initial state to a desired target state.\cite{MacKenzie2011}
Besides the use of fidelity to measure nonadiabaticity, two other
applications are explored here. First, we consider the importance
of additional terms in a nonadiabatic Hamiltonian, such as spin-orbit
coupling terms or couplings to an external field. In this case, both
$\hat{H}^{0}$ and $\hat{H}^{\epsilon}$ are coupled by nonadiabatic
coupling terms, and the additional term of interest, missing in $\hat{H}^{0}$,
is present in $\hat{H}^{\epsilon}$. Second, $F_{\mathrm{QM}}$ may
be used to evaluate quantitatively the accuracy of the quantum dynamics
with an approximate or interpolated nonadiabatic Hamiltonian $\hat{H}^{0}$
in comparison to the quantum dynamics with an accurate nonadiabatic
Hamiltonian $\hat{H}^{\epsilon}$. This application is a generalization
to nonadiabatic dynamics of the idea proposed for adiabatic dynamics
in Refs. \onlinecite{Li2009,Zimmermann2010a}.

The remaining (but difficult) question is how to compute $F_{\mathrm{QM}}$.
The most straightforward way would be to propagate the wave functions
with some wave packet method and to use Eq.\,\eqref{eq:Fidelity_definition}
directly. This approach, however, suffers from the previously mentioned
disadvantages of wave packet methods. Instead, below we derive an
accurate, yet efficient semiclassical method capable of computing
not only fidelity $F_{\mathrm{QM}}$ but also fidelity amplitude $f_{\mathrm{QM}}$.
The method, which we refer to as \textit{multiple-surface dephasing
representation} (MSDR), is a generalization of the dephasing representation
(DR),\cite{Vanicek2003,Vanicek2004,Vanicek2006} derived for the adiabatic
dynamics using the Van Vleck propagator. In the single surface setting,
the DR is closely related to the semiclassical perturbation approximation
of Hubbard and Miller\cite{Hubbard1983} and phase averaging of Mukamel.\cite{Mukamel1982}
Its main applications include calculations of electronic spectra\cite{Mukamel1982,Rost1995,Li1996,Egorov1998,Shi2005,Wehrle2011}
and evaluations of stability of quantum dynamics.\cite{Vanicek2004,Vanicek2006,Gorin2006,Wisniacki2010,Garcia-mata2011b}
The main advantage of the MSDR compared to wave packet methods is
that the computational cost of MSDR, similarly to the DR, does not
scale exponentially with the number of degrees of freedom.\cite{Mollica2011a}
The MSDR can therefore be applied to problems with dimensionality
far beyond the scope of current methods of quantum dynamics. The advantage
of MSDR in comparison to most other semiclassical approaches is that
the MSDR does not require the Hessian of the potential energy, which
is often the most expensive part of semiclassical calculations (see,
e.g., Ref. \onlinecite{Ceotto2009a}). Finally, in contrast to methods
treating the motion of nuclei completely classically, the MSDR includes
some nuclear quantum effects approximately via the interference between
the classical trajectories representing a wave packet. 

The MSDR is not the first extension of the DR to nonadiabatic dynamics.
In Ref. \onlinecite{Zimmermann2010} we have introduced another extension
of the DR, which is here referred to as \textit{integral multiple-surface
dephasing representation} (IMSDR) and which performs satisfactorily
in the case of nearly diabatic dynamics in the diabatic basis. Unfortunately,
the accuracy of the IMSDR deteriorates when the dynamics is far from
the diabatic limit. Another important limitation of the IMSDR is that
it cannot be used in the adiabatic basis. Below we shall demonstrate
that the MSDR is both more accurate and more general than the IMSDR.
A small price to pay for this improvement is that in contrast to the
IMSDR, in which fidelity is computed as an interference integral at
the end of dynamics, in the MSDR the Liouville-Von Neumann equation
for one discrete (collective electronic) degree of freedom has to
be solved during the dynamics. For pure states, this equation is simple
and is equivalent to the Schrödinger equation for one discrete degree
of freedom which is already solved during the Ehrenfest or fewest
switches surface hopping (FSSH) dynamics. Both MSDR and IMSDR reduce
to the original DR for systems with a single PES. 

The outline of the paper is as follows: in Section\,\ref{sec:Theory},
the MSDR is derived. In Section\,\ref{sec:Results}, the method is
used to evaluate nondiabaticity and nonadiabaticity of quantum dynamics
in model cases, to evaluate the importance of an additional coupling
term in a nonadiabatic Hamiltonian and to evaluate the accuracy of
an approximate Hamiltonian. Computational details are summarized in
the same section. Section\,\ref{sec:Discussion-and-Conclusions}
concludes the paper.

\section{Theory\label{sec:Theory}}

\subsection{MSDR}

A starting point for the derivation of the MSDR is the expression
for quantum fidelity amplitude formulated in terms of the density
matrix\cite{Vanicek2006} 
\begin{equation}
f_{\mathrm{QM}}\left(t\right)=\mathrm{Tr}\left(e^{-i\mathbf{\hat{H}}^{\epsilon}t/\hbar}\cdot\mathbf{\mathbf{\boldsymbol{\hat{\rho}}}}^{\text{init}}\cdot e^{+i\mathbf{\hat{H}}^{0}t/\hbar}\right),\label{eq:f_rho}
\end{equation}
 where $\boldsymbol{\hat{\rho}}^{\text{init}}$ is the density operator
of the initial state, $\mathbf{\hat{H}}^{0}$ and $\mathbf{\hat{H}}^{\epsilon}$
are two different molecular Hamiltonians expressed either in the diabatic
or adiabatic basis. (\textbf{Bold} face denotes $n\times n$ matrices
acting on the Hilbert space spanned by $n$ electronic states, hat
$\hat{}$ denotes nuclear operators.) Note that Eq. \eqref{eq:f_rho}
applies to both pure and mixed states.\cite{Vanicek2006} Formally,
$\mathbf{\hat{H}}^{\epsilon}$ can be written as $\mathbf{\hat{H}}^{\epsilon}=\mathbf{\hat{H}}^{0}+\epsilon\hat{\mathbf{V}}$,
where $\epsilon$ controls the extent of the perturbation. Expressing
$f_{\mathrm{QM}}$ in the interaction picture gives
\begin{equation}
f_{\mathrm{QM}}\left(t\right)=\mathrm{Tr}\left(\boldsymbol{\hat{\rho}}^{\text{init}}\cdot\hat{\mathbf{E}}(t)\right),\label{eq:f_int_pict}
\end{equation}
where
\begin{equation}
\hat{\mathbf{E}}\left(t\right):=e^{+i\mathbf{\hat{H}}^{0}t/\hbar}\cdot e^{-i\mathbf{\hat{H}}^{\epsilon}t/\hbar}={\cal T}e^{-i\epsilon\int_{0}^{t}\hat{\mathbf{V}}^{\text{I}}\left(t'\right)dt'/\hbar}\label{eq:echo_operator}
\end{equation}
is the echo operator\cite{Gorin2006,Veble2004} and
\begin{equation}
\mathbf{\hat{V}}^{\text{I}}\left(t\right)=e^{i\hat{\mathbf{H}}^{0}t/\hbar}\cdot\hat{\mathbf{V}}\cdot e^{-i\mathbf{\hat{H}}^{0}t/\hbar}\label{eq:pertubation_int_pict}
\end{equation}
is the perturbation in the interaction picture. A partial Wigner transform\cite{Wigner1932}
of Eq. \eqref{eq:f_int_pict} over nuclear degrees of freedom yields
an alternative exact expression for the fidelity amplitude, 
\begin{align}
f_{\mathrm{QM}}\left(t\right) & =h^{-D}\mathrm{Tr}_{e}\int dX\boldsymbol{\rho}_{\text{W}}^{\text{init}}\left(X\right)\cdot\mathbf{E}_{\text{W}}\left(X,t\right),\label{eq:f_int_pict_Wigner}
\end{align}
where $\mathbf{A}_{\text{W}}(X)$ is the partial Wigner transform
of operator $\hat{\mathbf{A}}$, 
\begin{equation}
\mathbf{A}_{\text{W}}(X)=\int d\xi\left\langle Q-\frac{\xi}{2}\right|\hat{\mathbf{A}}\left|Q+\frac{\xi}{2}\right\rangle \exp\left(i\frac{\xi\cdot P}{\hbar}\right),\label{eq:partial_Wigner}
\end{equation}
$X$ denotes the point $(Q,P)$ in the $2D$-dimensional phase space,
and $\mathrm{Tr}_{e}$ is the trace over electronic degrees of freedom.
Direct evaluation of the Wigner transform of the echo operator is
unfortunately difficult without approximations. As the first approximation,
one may truncate the Taylor expansion of the exponential operator
in the Wigner transform of a product of two operators 
\begin{eqnarray}
\left(\hat{\mathbf{A}}\cdot\hat{\mathbf{B}}\right)_{\text{W}} & = & \exp\left(\frac{i\hbar}{2}\left\{ .,.\right\} \right)\left(\mathbf{A}_{\text{W}},\mathbf{B}_{\text{W}}\right)\nonumber \\
 & = & \mathbf{A}_{\text{W}}\exp\left[\frac{i\hbar}{2}\left(\frac{\overleftarrow{\partial}}{\partial Q}\frac{\overrightarrow{\partial}}{\partial P}-\frac{\overleftarrow{\partial}}{\partial P}\frac{\overrightarrow{\partial}}{\partial Q}\right)\right]\mathbf{B}_{\text{W}}\label{eq:Wigner_product}
\end{eqnarray}
after the zeroth-order term. In Eq. \eqref{eq:Wigner_product}, $\left\{ \mathbf{A},\mathbf{B}\right\} =\frac{\partial\mathbf{A}}{\partial Q}\cdot\frac{\partial\mathbf{B}}{\partial P}-\frac{\partial\mathbf{A}}{\partial P}\cdot\frac{\partial\mathbf{B}}{\partial Q}$
is the Poisson bracket over the nuclear degrees of freedom and arrows
indicate on which argument the derivatives act. An iterative application
of expansion \eqref{eq:Wigner_product} to the echo operator \eqref{eq:echo_operator}
expressed as a time-ordered product of the short time propagators
gives 
\begin{equation}
\left({\cal T}e^{-i\epsilon\int_{0}^{t}\mathbf{\hat{V}}^{\text{I}}\left(t'\right)dt'/\hbar}\right)_{\text{W}}\simeq{\cal T}e^{-i\epsilon\int_{0}^{t}\mathbf{V}_{\text{W}}^{\text{I}}\left(X,t'\right)dt'/\hbar}.\label{eq:approx_product_Wigner_transf}
\end{equation}
To evaluate this expression, the time evolution of $\mathbf{V}_{\text{W}}^{\text{I}}\left(X,t\right)$
is required. The second approximation involves replacing the exact
equation of motion by a mixed quantum-classical (MQC) propagation
scheme described below, leading to the final expression for MSDR of
fidelity amplitude, 
\begin{eqnarray}
f_{\text{MSDR}}(t) & = & h^{-D}\mathrm{Tr}_{e}\int dX\boldsymbol{\rho}_{\text{W}}^{\text{init}}(X)\cdot{\cal T}e^{-i\epsilon\int_{0}^{t}\mathbf{V}_{\text{W,MQC}}^{\text{I}}\left(X,t'\right)dt'/\hbar}\nonumber \\
 & = & \left\langle {\cal T}e^{-i\epsilon\int_{0}^{t}\mathbf{V}_{\text{W,MQC}}^{\text{I}}\left(X,t'\right)dt'/\hbar}\right\rangle _{\boldsymbol{\rho}_{\text{W}}^{\text{init}}\left(X\right)},\label{eq:f_DR_D}
\end{eqnarray}
where the average in the last row is defined in general as
\[
\left\langle \mathbf{A}\left(X\right)\right\rangle _{\boldsymbol{\rho}\left(X\right)}:=\frac{\mathrm{Tr}_{e}\int dX\boldsymbol{\rho}(X)\cdot\mathbf{A}\left(X\right)}{\mathrm{Tr}_{e}\int dX\boldsymbol{\rho}(X)}.
\]
Equation\,\eqref{eq:f_DR_D} makes it clear that the only fundamental
difference between the MSDR and IMSDR introduced in Ref.\,\onlinecite{Zimmermann2010}
is the time ordering operator ${\cal T}$ present in the MSDR but
missing in the IMSDR.

\subsection{Propagation scheme}

Equation\,\eqref{eq:f_DR_D} gives $f_{\text{MSDR}}$ in terms of
$\mathbf{\mathbf{V}_{\text{W,MQC}}^{\text{I}}}(X,t)$; what remains
to be done is finding a prescription for $\mathbf{\mathbf{V}_{\text{W,MQC}}^{\text{I}}}(X,t)$.
The exact equation of motion for $\mathbf{V}_{\text{W}}^{\text{I}}\left(X,t\right)$
in the interaction picture is
\begin{equation}
\frac{\partial\mathbf{V}_{\text{W}}^{\text{I}}\left(X,t\right)}{\partial t}=\frac{i}{\hbar}\left[\hat{\mathbf{H}}^{0},\hat{\mathbf{V}}^{\text{I}}(t)\right]_{\text{W}}(X).\label{eq:operator_quantum_liouville_Wigner_transform}
\end{equation}
 Note that Eq.\,\eqref{eq:operator_quantum_liouville_Wigner_transform}
is, up to the sign, the same as the partially Wigner transformed Liouville-Von
Neumann equation describing the propagation of the density matrix
of the unperturbed system:
\begin{equation}
\frac{\partial\boldsymbol{\rho}_{\text{W}}(X,t)}{\partial t}=-\frac{i}{\hbar}\left[\hat{\mathbf{H}}^{0},\hat{\boldsymbol{\rho}}(t)\right]_{\text{W}}(X).\label{eq:quantum_liouville_Wigner_transform}
\end{equation}
We will take advantage of this analogy and derive our approximate
propagation scheme from Eq.\,\eqref{eq:quantum_liouville_Wigner_transform}
instead of from Eq.\,\eqref{eq:operator_quantum_liouville_Wigner_transform}.
This will turn out to be slightly easier and, more importantly, we
will simultaneously find an approximate solution of a much more general
problem of propagating the density matrix. Below we omit superscript
0 in $\hat{\mathbf{H}}^{0}$ since now we deal with a single Hamiltonian.
In the system consisting of light electrons and heavy nuclei, Eq.\,\eqref{eq:quantum_liouville_Wigner_transform}
is approximated to the first order in the mass ratio $\sqrt{\frac{m}{M}}$
by the mixed quantum-classical Liouville equation\cite{Aleksandrov1981,Boucher1988,Martens1997,Prezhdo1997,Kapral1999,Caro1999,Shi2004}
\begin{align}
\frac{\partial\boldsymbol{\rho}_{\text{W,MQC}}}{\partial t}= & -\frac{i}{\hbar}\left[\mathbf{H}_{\text{W}},\boldsymbol{\rho}_{\text{W,MQC}}\right]
\nonumber\\ & +\frac{1}{2}\left(\left\{ \mathbf{H}_{\text{W}},\boldsymbol{\rho}_{\text{W,MQC}}\right\} -\left\{ \boldsymbol{\rho}_{\text{W,MQC}},\mathbf{H}_{\mathrm{\text{W}}}\right\} \right),\label{eq:mixed_Liouville}
\end{align}
where the explicit dependence of $\boldsymbol{\rho}_{\text{W,MQC}}$
on time and on the nuclear phase space coordinate $X$ was omitted
for clarity. 

Equation \eqref{eq:f_DR_D} together with Eq. \eqref{eq:mixed_Liouville}
define the MSDR. Several numerical approaches exist that solve Eq.\,\eqref{eq:mixed_Liouville}
in terms of ``classical'' trajectories $X(t)$. However, since trajectory
based methods for solving Eq.\,\eqref{eq:mixed_Liouville} are still
relatively complicated, below we study two variants of MSDR where
Eq. \eqref{eq:mixed_Liouville} is further approximated. The common
feature of the two approximations is that all elements of $\boldsymbol{\rho}_{\text{W}}\left(X,t\right)$
are propagated using the same PES (which may, nevertheless, differ
for different trajectories). For simplicity, from now on the subscript
MQC is omitted.

\subsubsection{Locally mean field dynamics}

The first approach starts by rewriting $\boldsymbol{\rho}_{\text{W}}\left(X,t\right)$
as 
\begin{equation}
\boldsymbol{\rho}_{\text{W}}(X,t)=\rho(X,t)\boldsymbol{\rho}_{e}(X,t),\label{eq:rho_LMFD}
\end{equation}
where $\rho(X,t):=\mathrm{Tr}_{e}\boldsymbol{\rho}_{\text{W}}(X,t)$
is a scalar function of $X$ and $t\mbox{ and }$hence $\mathrm{Tr}_{e}\boldsymbol{\rho}_{e}(X,t)=1$
for all $X$ and $t$. By substituting the still exact ansatz \eqref{eq:rho_LMFD}
into Eq.\,\eqref{eq:mixed_Liouville}, one obtains
\begin{align}
\frac{\partial\rho}{\partial t}\boldsymbol{\rho}_{e}+\rho\frac{\partial\boldsymbol{\rho}_{e}}{\partial t}= & -\frac{i}{\hbar}\rho\left[\mathbf{H}_{\text{W}},\boldsymbol{\rho}_{e}\right]\nonumber \\
 & +\frac{1}{2}\frac{\partial\rho}{\partial P}\left[\frac{\partial\mathbf{H}_{\text{W}}}{\partial Q},\boldsymbol{\rho}_{e}\right]_{+}\nonumber \\
 & +\frac{1}{2}\rho\left[\frac{\partial\mathbf{H}_{\text{W}}}{\partial Q},\frac{\partial\boldsymbol{\rho}_{e}}{\partial P}\right]_{+}\nonumber \\
 & -\frac{1}{2}\frac{\partial\rho}{\partial Q}\left[\frac{\partial\mathbf{H}_{\text{W}}}{\partial P},\boldsymbol{\rho}_{e}\right]_{+}\nonumber \\
 & -\frac{1}{2}\rho\left[\frac{\partial\mathbf{H}_{\text{W}}}{\partial P},\frac{\partial\boldsymbol{\rho}_{e}}{\partial Q}\right]_{+},\label{eq:mixed_Liouville_with_ansatz}
\end{align}
where $\left[\mathbf{A},\mathbf{B}\right]_{+}=\mathbf{A}\cdot\mathbf{B}+\mathbf{B}\cdot\mathbf{A}$
is the anticommutator. In the next step, the trace over the electronic
degrees of freedom is performed, which in the diabatic basis leads
to the following equation of motion for $\rho\left(X,t\right)$:

\begin{align}
\frac{\partial\rho}{\partial t} & =\frac{\partial\rho}{\partial P}\left\langle \frac{\partial\mathbf{H}_{\text{W}}}{\partial Q}\right\rangle _{e}-\frac{\partial\rho}{\partial Q}\left\langle \frac{\partial\mathbf{H}_{\text{W}}}{\partial P}\right\rangle _{e}\nonumber \\
 & +\rho\mathrm{Tr}_{e}\left(\frac{\partial\mathbf{H}_{\text{W}}}{\partial Q}\cdot\frac{\partial\boldsymbol{\rho}_{e}}{\partial P}\right),\label{eq:nuclear_Liouville_diab_full}
\end{align}
where $\left\langle \mathbf{A}\right\rangle _{e}=\mathrm{Tr}_{e}\left(\boldsymbol{\rho}_{e}\cdot\mathbf{A}\right)$
is a partial average of $\mathbf{A}$ over the electronic subspace
and where we have used that $\mathrm{Tr}_{e}\left(\frac{\partial\mathbf{H}_{\text{W}}}{\partial P}\cdot\frac{\partial\boldsymbol{\rho}_{e}}{\partial Q}\right)=\frac{P}{M}\mathrm{Tr}_{e}\left(\frac{\partial\boldsymbol{\rho}_{e}}{\partial Q}\right)=0$.
(The equation of motion in the adiabatic basis will be derived later.)
Substitution of Eq. \eqref{eq:nuclear_Liouville_diab_full} back into
Eq.\,\eqref{eq:mixed_Liouville_with_ansatz} yields
\begin{eqnarray}
\rho\frac{\partial\boldsymbol{\rho}_{e}}{\partial t}= & - & \frac{i}{\hbar}\rho\left[\mathbf{H}_{\text{W}},\boldsymbol{\rho}_{e}\right]\nonumber \\
 & + & \frac{\partial\rho}{\partial P}\left(\frac{1}{2}\left[\frac{\partial\mathbf{H}_{\text{W}}}{\partial Q},\boldsymbol{\rho}_{e}\right]_{+}-\left\langle \frac{\partial\mathbf{H}_{\text{W}}}{\partial Q}\right\rangle _{e}\cdot\boldsymbol{\rho}_{e}\right)\nonumber \\
 & + & \frac{1}{2}\rho\left[\frac{\partial\mathbf{H}_{\text{W}}}{\partial Q},\frac{\partial\boldsymbol{\rho}_{e}}{\partial P}\right]_{+}\nonumber \\
 & - & \frac{\partial\rho}{\partial Q}\left(\frac{1}{2}\left[\frac{\partial\mathbf{H}_{\text{W}}}{\partial P},\boldsymbol{\rho}_{e}\right]_{+}-\left\langle \frac{\partial\mathbf{H}_{\text{W}}}{\partial P}\right\rangle _{e}\cdot\boldsymbol{\rho}_{e}\right)\nonumber \\
 & - & \rho\left\langle \frac{\partial\mathbf{H}_{\text{W}}}{\partial P}\right\rangle _{e}\cdot\frac{\partial\boldsymbol{\rho}_{e}}{\partial Q}\nonumber \\
 & - & \rho\boldsymbol{\rho}_{e}\mathrm{Tr}\left(\frac{\partial\mathbf{H}_{\text{W}}}{\partial Q}\cdot\frac{\partial\boldsymbol{\rho}_{e}}{\partial P}\right),\label{eq:electronic_Liouville_diab_full}
\end{eqnarray}
where identity $\frac{1}{2}\left[\frac{\partial\mathbf{H}_{\text{W}}}{\partial P},\frac{\partial\boldsymbol{\rho}_{e}}{\partial Q}\right]_{+}=\frac{P}{M}\frac{\partial\boldsymbol{\rho}_{e}}{\partial Q}=\left\langle \frac{\partial\mathbf{H}_{\text{W}}}{\partial P}\right\rangle _{e}\cdot\frac{\partial\boldsymbol{\rho}_{e}}{\partial Q}$
was used in the fifth row. Both Eqs.\,\eqref{eq:nuclear_Liouville_diab_full}
and \eqref{eq:electronic_Liouville_diab_full} contain terms of the
form $\left\langle \frac{\partial\mathbf{H}_{\text{W}}}{\partial P}\right\rangle _{e}\frac{\partial f}{\partial Q}$
and $\left\langle \frac{\partial\mathbf{H}_{\text{W}}}{\partial Q}\right\rangle _{e}\frac{\partial f}{\partial P}$,
which can be combined with the time derivative $\frac{\partial f}{\partial t}$
to form the convective derivative
\[
\frac{Df}{Dt}=\frac{\partial f}{\partial t}+\dot{Q}\frac{\partial f}{\partial Q}+\dot{P}\frac{\partial f}{\partial P}
\]
 and transform the equations from the Eulerian reference frame at
rest to the Lagrangian reference frame moving with the phase space
flow given by the vector field
\begin{equation}
\left(\dot{Q},\dot{P}\right)=\left(\left\langle \frac{\partial\mathbf{H}_{\text{W}}}{\partial P}\right\rangle _{e},-\left\langle \frac{\partial\mathbf{H}_{\text{W}}}{\partial Q}\right\rangle _{e}\right).\label{eq:vector_field}
\end{equation}
 In the Lagrangian frame, Eq.\,\eqref{eq:nuclear_Liouville_diab_full}
transforms to 
\begin{equation}
\frac{D\rho}{Dt}=\rho\mathrm{Tr}_{e}\left(\frac{\partial\mathbf{H}_{\text{W}}}{\partial Q}\cdot\frac{\partial\boldsymbol{\rho}_{e}}{\partial P}\right).\label{eq:mixed_Liouville_Lagrangian_nuclei}
\end{equation}
Since the two terms in the fourth row of Eq.\,\eqref{eq:electronic_Liouville_diab_full}
cancel each other exactly, this equation transforms to 
\begin{align}
\frac{D\boldsymbol{\rho}_{e}}{Dt}= & -\frac{i}{\hbar}\left[\mathbf{H}_{W},\boldsymbol{\rho}_{e}\right]\nonumber \\
 & +\frac{1}{\rho}\frac{\partial\rho}{\partial P}\left(\frac{1}{2}\left[\frac{\partial\mathbf{H}_{W}}{\partial Q},\boldsymbol{\rho}_{e}\right]_{+}-\left\langle \frac{\partial\mathbf{H}_{W}}{\partial Q}\right\rangle _{e}\boldsymbol{\rho}_{e}\right)\nonumber \\
 & +\left(\frac{1}{2}\left[\frac{\partial\mathbf{H}_{W}}{\partial Q},\frac{\partial\boldsymbol{\rho}_{e}}{\partial P}\right]_{+}-\left\langle \frac{\partial\mathbf{H}_{\text{W}}}{\partial Q}\right\rangle _{e}\cdot\frac{\partial\boldsymbol{\rho}_{e}}{\partial P}\right)\nonumber \\
 & -\boldsymbol{\rho}_{e}\mathrm{Tr}_{e}\left(\frac{\partial\mathbf{H}_{\text{W}}}{\partial Q}\cdot\frac{\partial\boldsymbol{\rho}_{e}}{\partial P}\right).\label{eq:mixed_Liouville_Lagrangian_electrons}
\end{align}
 Note that the second and third row of the right hand side of Eq.\,\eqref{eq:mixed_Liouville_Lagrangian_electrons}
contain differences (put in parentheses for emphasis) between products
of the electronic density matrix (or its $P$ derivative) with the
nonaveraged and averaged gradients of the Hamiltonian. Therefore,
the second and third tow may often be small compared to the first
and fourth row of Eq.\,\eqref{eq:mixed_Liouville_Lagrangian_electrons}.
The last term on the right hand side of Eq.\,\eqref{eq:mixed_Liouville_Lagrangian_electrons}
corresponds to the right hand side of Eq.\,\eqref{eq:mixed_Liouville_Lagrangian_nuclei}
and is responsible for maintaining $\mathrm{Tr}_{e}\boldsymbol{\rho}_{e}(X,t)=1$
during the (non-approximated) MQC dynamics. Until now all operations
have been exact and Eqs.\,\eqref{eq:mixed_Liouville_Lagrangian_nuclei}
and \eqref{eq:mixed_Liouville_Lagrangian_electrons} are equivalent
to the original mixed quantum classical Liouville equation\,\eqref{eq:mixed_Liouville}.

Now we will make the locally mean field approximation: Specifically,
once in the Lagrangian reference frame, all terms in both Eqs.\,\eqref{eq:mixed_Liouville_Lagrangian_nuclei}
and \eqref{eq:mixed_Liouville_Lagrangian_electrons} containing phase
space derivatives of $\rho\left(X,t\right)$ or $\boldsymbol{\rho}_{e}\left(X,t\right)$
are neglected to obtain the approximate locally mean-field equations
of motion. The resulting equation for $\rho\left(X,t\right)$ in the
Lagrangian reference frame is simply 
\begin{equation}
\frac{D\rho_{\mathrm{LMFD}}}{Dt}=0,\label{eq:nuclear_Liouville_diab_local}
\end{equation}
and the equation of motion for the electronic part of the density
matrix $\boldsymbol{\rho}_{e}\left(X,t\right)$ is 
\begin{align}
\frac{D\boldsymbol{\rho}_{e,\mathrm{LMFD}}}{Dt}= & -\frac{i}{\hbar}\left[\mathbf{H}_{\text{W}},\boldsymbol{\rho}_{e,\mathrm{LMFD}}\right].\label{eq:electronic_Liouville_diab_local}
\end{align}
Both equations can be combined together and transformed back to the
Eulerian reference frame to yield the equation of motion for the total
density matrix $\boldsymbol{\rho}_{\text{W}}(X,t)$, 
\begin{eqnarray}
\frac{\partial\boldsymbol{\rho}_{\text{W,LMFD}}}{\partial t} & = & -\frac{i}{\hbar}\left[\mathbf{H}_{\text{W}},\boldsymbol{\rho}_{\text{W,LMFD}}\right]\nonumber \\
 &  & +\frac{\partial\boldsymbol{\rho}_{\text{W,LMFD}}}{\partial P}\left\langle \frac{\partial\mathbf{H}_{\text{W}}}{\partial Q}\right\rangle _{e} \nonumber \\
 &  & -\frac{\partial\boldsymbol{\rho}_{\text{W,LMFD}}}{\partial Q}\frac{P}{M},\label{eq:Liouville_diab_local}
\end{eqnarray}
 where we have used that $\left\langle \frac{\partial\mathbf{H}_{\text{W}}}{\partial P}\right\rangle _{e}=\frac{P}{M}$.

We call the dynamics expressed by Eq.\,\eqref{eq:Liouville_diab_local}
{[}or, equivalently, by Eqs.\,\eqref{eq:vector_field}, \eqref{eq:nuclear_Liouville_diab_local}
and \eqref{eq:electronic_Liouville_diab_local}{]} the \textit{locally
mean field dynamics} (LMFD) since the force acting on nuclei at position
$X$ is the force averaged over the ``electronic'' part of the density
matrix $\boldsymbol{\rho}_{e}(X,t)$ at $X$. Note that the well-known
\textit{Ehrenfest dynamics} (which uses a single nuclear trajectory)
can be derived in a similar way using an approximate ansatz $\boldsymbol{\rho}_{\text{W}}^{\text{ED}}=\delta(X-X\left(t\right))\boldsymbol{\rho}_{e}\left(t\right)$,
where $\delta$ is the Dirac delta distribution and $X\left(t\right)$
is the phase space coordinate at time $t$.\cite{Grunwald2009} Similarly,
a truly \textit{mean field dynamics} for a wave packet different from
a $\delta$ distribution can be derived using an (again approximate)
ansatz in the form of a Hartree product $\boldsymbol{\rho}_{\text{W}}^{\text{MFD}}\left(X,t\right)=\rho(X,t)\boldsymbol{\rho}_{e}\left(t\right)$. 

As can be seen from Eq.\,\eqref{eq:Liouville_diab_local}, for pure
states, each phase space point is propagated by the mean field Ehrenfest
dynamics, according to Eq. \eqref{eq:vector_field}. Nevertheless,
in contrast to the purely mean field dynamics, Eq.\,\eqref{eq:Liouville_diab_local}
describes the propagation of the density $\boldsymbol{\rho}_{\text{W}}(X,t)$
using different values of $\left\langle \frac{\partial\mathbf{H}_{\text{W}}}{\partial Q}\right\rangle _{e}$
and $\left\langle \frac{\partial\mathbf{H}_{\text{W}}}{\partial P}\right\rangle _{e}$
for each value of $X$. Interestingly, this dynamics corresponds exactly
to the nuclear dynamics appearing in the nonadiabatic IVR model,\cite{Ananth2007,Miller2009}
which uses the Meyer-Miller-Stock-Thoss Hamiltonian.\cite{Meyer1979,Stock1997}
Also note that outside of coupling regions and when only one surface
is occupied, resulting trajectories are equivalent to those obtained
with the Born-Oppenheimer dynamics. 

To derive a corresponding LMFD equation of motion in the adiabatic
basis, one can express the mixed quantum classical Liouville equation
\eqref{eq:mixed_Liouville} in the adiabatic basis,\cite{Kapral1999}
use the exact ansatz \eqref{eq:rho_LMFD}, and apply a similar LMFD
approximation as above. However, this procedure is quite tedious in
the adiabatic basis. The LMFD in adiabatic basis can be obtained more
easily by directly transforming the final LMFD equation of motion
\eqref{eq:nuclear_Liouville_diab_local} from diabatic to adiabatic
basis using the relations\cite{Kapral1999} 
\begin{eqnarray}
\left(\frac{\partial\mathbf{A}_{\text{W}}}{\partial Q}\right)^{\mathrm{A}} & = & \frac{\partial\mathbf{A}_{\text{W}}^{\text{A}}}{\partial Q}-\left[\mathbf{A}_{\text{W}}^{\text{A}},\mathbf{F}\right]\;\text{and}\label{eq:d_rho_adiab}\\
\left(\frac{\partial\mathbf{A}_{\text{W}}}{\partial P}\right)^{\mathrm{A}} & = & \frac{\partial\mathbf{A}_{\text{W}}^{\text{A}}}{\partial P},\label{eq:d_rho_adiab-P}
\end{eqnarray}
where the superscript A denotes the transformation to the adiabatic
basis and $\mathbf{F}$ is the vector matrix of nonadiabatic coupling
vectors. Specifically, $\mathbf{A}_{\text{W}}^{\text{A}}(X,t)$ is
a matrix obtained by first partially Wigner transforming a general
operator $\hat{\mathbf{A}}(t)$ {[}to form $\mathbf{A}_{\text{W}}\left(X,t\right)${]}
and then by transforming $\mathbf{A}_{\text{W}}\left(X,t\right)$
into the adiabatic basis. Matrix $\mathbf{F}$ is defined componentwise
by $F_{jk}\left(Q\right)=\langle\alpha_{j}\left(Q\right)\vert\frac{\partial}{\partial Q}\alpha_{k}\left(Q\right)\rangle$,
where $\vert\alpha_{k}\left(Q\right)\rangle$ is $k$-th element of
the adiabatic basis at position $Q$. Applying relations \eqref{eq:d_rho_adiab}
and \eqref{eq:d_rho_adiab-P} to derivatives of both $\mathbf{H}_{\text{W}}$
and $\boldsymbol{\rho}_{\text{W,LMFD}}$ in Eq. \eqref{eq:Liouville_diab_local}
immediately yields the LMFD equation of motion in the adiabatic basis,
\begin{eqnarray}
\frac{\partial\boldsymbol{\rho}_{\text{W,LMFD}}^{\text{A}}}{\partial t} & = & -\frac{i}{\hbar}\left[\mathbf{H}_{\text{W}}^{\text{A}}(X)-i\hbar\frac{P}{M}\mathbf{F}(Q),\boldsymbol{\rho}_{\text{W,LMFD}}^{\text{A}}\right]\nonumber \\
 &  & +\frac{\partial\boldsymbol{\rho}_{\text{W,LMFD}}^{\text{A}}}{\partial P}\cdot\left(\left\langle \frac{\partial\mathbf{H}_{\text{W}}^{\text{A}}}{\partial Q}\right\rangle _{e}-\left\langle \left[\mathbf{H}_{\text{W}}^{\text{A}},\mathbf{F}\right]\right\rangle _{e}\right)\nonumber \\
 &  & -\frac{\partial\boldsymbol{\rho}_{\text{W,LMFD}}^{\text{A}}}{\partial Q}\frac{P}{M},\label{eq:Liouville_adiab_1_local}
\end{eqnarray}
where $\mathbf{H}_{\text{W}}^{\text{A}}$ is the diabatic Hamiltonian
matrix $\mathbf{H}_{\text{W}}$ expressed in the adiabatic basis.
Note that $\mathbf{H}_{\text{W}}^{\text{A}}$ consists only of the
kinetic energy term and the diagonal adiabatic potential energy matrix;
in particular, $\mathbf{H}_{\text{W}}^{\text{A}}$ does not contain
the nonadiabatic couplings. Again, the dynamics of a single trajectory
is identical to the Ehrenfest dynamics.

\subsubsection{Fewest switches surface hopping}

The second alternative approximate propagation scheme is based on
the physically motivated FSSH algorithm.\cite{Tully1990} In this
scheme, each phase space point representing the Wigner density distribution
is propagated independently using the FSSH dynamics. Compared to the
LMFD, the FSSH is used at no additional cost, except for the stochastic
hopping algorithm itself, because the same Eq.\,\eqref{eq:electronic_Liouville_diab_local}
(or its equivalent in the adiabatic basis) has to be solved during
the electronic part of the dynamics. On the other hand, averaging
the force over electronic states is avoided in the FSSH. As will be
shown below, neither method is universally optimal and the best propagation
method depends on a problem studied.

\subsection{Algorithm}

\subsubsection{General initial state}

To compute $f_{\text{MSDR}}(t)$, we rewrite the initial density matrix
exactly in the form 
\begin{equation}
\boldsymbol{\rho}_{\text{W}}^{\text{init}}\left(X\right)=\rho^{\text{init}}\left(X\right)\boldsymbol{\rho}_{e}^{\text{init}}\left(X\right),\label{eq:Initial density matrix}
\end{equation}
where $\rho^{\text{init}}\left(X\right):=\mathrm{Tr}_{e}\boldsymbol{\rho}_{\mathrm{W}}^{\text{init}}\left(X\right)$.
Scalar nuclear density $\rho^{\text{init}}(X)$ is sampled by $N_{\text{traj}}$
phase space points that serve as initial conditions for $N_{\text{traj}}$
trajectories propagated using either the LMFD or FSSH dynamics. For
each generated phase space point, the electronic part $\boldsymbol{\rho}_{e}^{\text{init}}\left(X\right)$
satisfies $\mathrm{Tr}_{e}\boldsymbol{\rho}_{e}^{\text{init}}\left(X\right)=1$.
In the case of LMFD, $X$ determines the initial condition completely.
In the case of FSSH, one also needs to select the initial surface
randomly for each trajectory according to the following prescription:
For a trajectory starting at $X$, the probability for its initial
surface to be surface $j$ is given by the diagonal element $\rho_{e,jj}^{\text{init}}(X)$.
Equation\,\eqref{eq:f_DR_D} for fidelity amplitude is rewritten
as
\begin{align}
& f_{\text{MSDR}}\left(t\right)= \nonumber \\
& \left\langle \mathrm{Tr}_{e}\left[\boldsymbol{\rho}_{e}^{\text{init}}\left(X\right){\cal \cdot T}e^{-i\epsilon\int_{0}^{t}\mathbf{V}_{\text{W}}^{\text{I}}\left(X,t'\right)dt'/\hbar}\right]\right\rangle _{\rho^{\text{init}}\left(X\right)},\label{eq:f_DR_D_explicit}
\end{align}
where we have used the notation 
\[
\left\langle A\left(X\right)\right\rangle _{\rho\left(X\right)}:=\frac{\int dX\rho(X)A\left(X\right)}{\int dX\rho(X)}
\]
and the fact that $\left\langle \mathrm{Tr}_{e}\boldsymbol{\rho}_{e}^{\text{init}}\left(X\right)\right\rangle _{\rho^{\text{init}}\left(X\right)}=1.$
The time-ordered product ${\cal T}e^{-i\epsilon\int_{0}^{t}\mathbf{V}_{\text{W}}^{\text{I}}\left(X,t'\right)dt'/\hbar}$
is evaluated along each trajectory by successive multiplication of
short time propagators corresponding to each time step. The exponent
of each short time propagator is computed using
\begin{equation}
\mathbf{V}_{\text{W}}^{\text{I}}(X,t)=\mathbf{U}^{0}(X,t)^{-1}\cdot\mathbf{V}_{\text{W}}(X^{0}(t))\cdot\mathbf{U}^{0}(X,t),\label{eq:v_int_explicit}
\end{equation}
where 
\begin{equation}
\mathbf{U}^{0}(X,t)={\cal T}e^{-i\intop_{0}^{t}\mathbf{H}_{\text{W}}^{0}\left(X^{0}(t')\right)dt'/\hbar},\label{eq:U_0}
\end{equation}
and $X^{0}(t)$ denotes a trajectory of $\mathbf{H}_{\text{W}}^{0}$
starting at $X^{0}(0)=X$. Equation \eqref{eq:v_int_explicit} can
be vaguely interpreted as a combination of a quantum interaction picture
for the electrons and a ``classical interaction picture'' (in which
the perturbation is neglected) for the nuclei. Operator $\mathbf{U}^{0}(X,t)$
is again computed by successive multiplication of short time propagators
along the trajectory. At the end of the dynamics, the weighted phase
space average in Eq.\,\eqref{eq:f_DR_D_explicit} is computed as
the arithmetic average over all trajectories:
\begin{align}
& f_{\text{MSDR}}\left(t\right)= \nonumber \\
& \frac{1}{N_{\text{traj}}}\sum_{N=1}^{N_{\text{traj}}}\mathrm{Tr}_{e}\left[\boldsymbol{\rho}_{e}^{\text{init}}\left(X_{N}\right){\cal \cdot T}\prod_{M=1}^{M{}_{\text{step}}}e^{-i\epsilon\mathbf{V}_{\text{W}}^{\text{I}}\left(X_{N},M\Delta t\right)\Delta t/\hbar}\right].\label{eq:numerical_DR_D}
\end{align}

\subsubsection{Initial Hartree product state}

The algorithm simplifies slightly when the initial density operator
$\hat{\boldsymbol{\rho}}^{\text{init}}$ is a Hartree product 
\begin{equation}
\hat{\boldsymbol{\rho}}^{\text{init}}=\hat{\rho}_{N}^{\text{init}}\otimes\boldsymbol{\rho}_{e}^{\text{init}},\label{eq:Initial_hartree_product}
\end{equation}
where $\hat{\rho}_{N}^{\text{init}}$ and $\boldsymbol{\rho}_{e}^{\text{init}}$
are the nuclear and electronic density operators, respectively. This
Hartree approximation is usually an excellent approximation outside
of coupling regions when only one PES is occupied. For initial density
in the product form \eqref{eq:Initial_hartree_product}, $\boldsymbol{\rho}_{\text{W}}^{\text{init}}\left(X\right)=\rho^{\text{init}}\left(X\right)\boldsymbol{\rho}_{e}^{\text{init}}$,
where $\rho^{\text{init}}\left(X\right)=\left[\rho_{N}^{\text{init}}\right]_{\text{W}}\left(X\right)$
and $\boldsymbol{\rho}_{e}^{\text{init}}$ is independent of $X$.
Equations \eqref{eq:f_DR_D_explicit} and \eqref{eq:numerical_DR_D}
become
\begin{align}
& f_{\text{MSDR}}\left(t\right)= \nonumber \\
& \mathrm{Tr}_{e}\left[\boldsymbol{\rho}_{e}^{\text{init}}\cdot\left\langle {\cal T}e^{-i\epsilon\int_{0}^{t}\mathbf{V}_{\text{W}}^{\text{I}}\left(X,t'\right)dt'/\hbar}\right\rangle _{\rho^{\text{init}}\left(X\right)}\right]
\end{align}
and
\begin{align}
& f_{\text{MSDR}}\left(t\right)= \nonumber \\
& \mathrm{Tr}_{e}\left[\boldsymbol{\rho}_{e}^{\text{init}}\cdot\frac{1}{N_{\text{traj}}}\sum_{N=1}^{N_{\text{traj}}}{\cal T}\prod_{M=1}^{M{}_{\text{step}}}e^{-i\epsilon\mathbf{V}_{\text{W}}^{\text{I}}\left(X_{N},M\Delta t\right)\Delta t/\hbar}\right].
\end{align}

\subsubsection{``Electronically pure'' initial state}

The algorithm simplifies also for ``electronically pure'' initial
states, by which we mean (in the most general sense) states for which
the electronic density matrix $\boldsymbol{\rho}_{e}^{\text{init}}\left(X\right)$
in the product \eqref{eq:Initial density matrix} is pure for all
$X$, i.e., satisfies the condition $\mathrm{Tr}_{e}\left[\boldsymbol{\rho}_{e}^{\text{init}}\left(X\right)^{2}\right]=1$
and can be written as the tensor product
\begin{equation}
\boldsymbol{\rho}_{e}^{\text{init}}\left(X\right)=\mathbf{c^{\mathrm{init}}}\left(X\right)\otimes\mathbf{c}^{\mathrm{init}}\left(X\right)^{\dagger}\label{eq:init_density_pure}
\end{equation}
in terms of an initial electronic wave function $\mathbf{c^{\mathrm{init}}}\left(X\right)$.
The generalized electronically pure states include, as a special case,
the Hartree product \eqref{eq:Initial_hartree_product} in which the
constant electronic matrix $\boldsymbol{\rho}_{e}^{\text{init}}$
is pure (while the nuclear factor $\hat{\rho}_{N}^{\text{init}}$
may be mixed). To derive the simplified expression for $f_{\text{MSDR}}\left(t\right)$,
we first rewrite the approximate Wigner transform of the echo operator
in Eq. \eqref{eq:f_DR_D_explicit} as a product of the electronic
evolution operators: 
\begin{equation}
{\cal T}e^{-i\epsilon\int_{0}^{t}\mathbf{V}_{\text{W}}^{\text{I}}\left(X,t'\right)dt'/\hbar}=\mathbf{U}^{0}\left(X,t\right)^{-1}\cdot\mathbf{U}^{\epsilon}(X,t),\label{eq:wigner_echo_lagrangian-1}
\end{equation}
where $\mathbf{U}^{0}(X,t)$ was defined in Eq. \eqref{eq:U_0} and,
similarly,
\begin{equation}
\mathbf{U}^{\epsilon}(X,t)={\cal T}e^{-i\intop_{0}^{t}\mathbf{H}_{\text{W}}^{\epsilon}\left(X^{0}(t')\right)dt'/\hbar}.\label{eq:U_eps}
\end{equation}
Note that the electronic evolution operators $\mathbf{U}^{0}\left(X,t\right)$
and $\mathbf{U}^{\epsilon}(X,t)$ are defined separately for each
nuclear trajectory. Using expression \eqref{eq:wigner_echo_lagrangian-1},
we can rewrite the MSDR of fidelity amplitude \eqref{eq:f_DR_D_explicit}
as 
\begin{align}
& f_{\text{MSDR}}\left(t\right)= \nonumber \\
& \left\langle \mathrm{Tr}_{e}\left[\boldsymbol{\rho}_{e}^{\text{init}}\left(X\right)\cdot\mathbf{U}^{0}\left(X,t\right)^{-1}\cdot\mathbf{U}^{\epsilon}(X,t)\right]\right\rangle _{\rho^{\text{init}}\left(X\right)}.\label{eq:f_DR_D_alternative-1}
\end{align}
This expression, which is still valid for any initial state, is equivalent
to Eq.\,\eqref{eq:f_DR_D_explicit} when either the LMFD or FSSH
dynamics is used for propagation.

For electronically pure states \eqref{eq:init_density_pure}, Eq.\,\eqref{eq:f_DR_D_alternative-1}
further simplifies into the weighted phase space average 
\begin{equation}
f_{\text{MSDR}}\left(t\right)=\left\langle \mathbf{c}^{0}\left(X,t\right)^{\dagger}\cdot\mathbf{c}^{\epsilon}\left(X,t\right)\right\rangle _{\rho^{\text{init}}\left(X\right)},\label{eq:f_DR_D_pure_states-1}
\end{equation}
where $\mathbf{c}^{\epsilon}\left(X,t\right)$ is the wave function
with initial condition $\mathbf{c}^{\epsilon}\left(X,0\right)=\mathbf{c}^{\mathrm{init}}\left(X\right)$
that solves the Schr\"odinger equation $\frac{\partial\mathbf{c}^{\epsilon}\left(X,t\right)}{\partial t}=-\frac{i}{\hbar}\mathbf{H}_{\text{W}}^{\epsilon}(X^{0}(t))\cdot\mathbf{c}^{\epsilon}\left(X,t\right)$
for a single discrete electronic degree of freedom in the Lagrangian
reference frame given by $\mathbf{H}_{\text{W}}^{0}$. In both propagation
algorithms currently used with the MSDR, i.e., the LMFD and FSSH dynamics,
$\mathbf{c}^{0}\left(X,t\right)$ is already available; only $\mathbf{c}^{\epsilon}\left(X,t\right)$
has to be determined additionally. Expressed explicitly in terms of
trajectories, Eq.\,\eqref{eq:f_DR_D_pure_states-1} allows computing
the fidelity amplitude simply as

\begin{equation}
f_{\text{MSDR}}\left(t\right)=\frac{1}{N_{\text{traj}}}\sum_{N=1}^{N_{\text{traj}}}\mathbf{c}_{N}^{0}\left(X,t\right)^{\dagger}\cdot\mathbf{c}_{N}^{\epsilon}\left(X,t\right).\label{eq:numerical_DR_D_pure_states}
\end{equation}

\section{Results\label{sec:Results}}

To test the MSDR, we considered several model systems allowing comparison
with the exact quantum-mechanical solution in both the diabatic and
adiabatic bases. Specifically, we used variants of the three one-dimensional
model potentials introduced by Tully\cite{Tully1990} and the simple
two-level model of photodissociation of NaI.\cite{Engel1989,Veen1981,Faist1976}
The mass in Tully's models was set to 2000 $\mathrm{a.u.}$, approximately
corresponding to the mass of the hydrogen atom. The reduced mass in
the model of photodissociation of NaI was set to 35480.25 $\mathrm{a.u.}$ 

The initial density matrix was in all cases a density matrix of a
pure state, so $f_{\text{MSDR}}$ was evaluated using Eq.\,\eqref{eq:numerical_DR_D_pure_states}.
In Subsections\,\ref{sub:Diabatic-basis} and \ref{sub:Adiabatic-basis},
fidelity is used as a quantitative measure of the importance of the
diabatic or nonadiabatic couplings on the dynamics. In other words,
Hamiltonian $\mathbf{\hat{H}}^{0}$ is the diagonal diabatic (Subsec.\,\ref{sub:Diabatic-basis})
or adiabatic (Subsec.\,\ref{sub:Adiabatic-basis}) Hamiltonian and
Hamiltonian $\mathbf{\hat{H}}^{\epsilon}$ is the full nondiabatic
(Subsec.\,\ref{sub:Diabatic-basis}) or nonadiabatic (Subsec.\,\ref{sub:Adiabatic-basis})
Hamiltonian, respectively. If not mentioned otherwise, the dynamics
on $\mathbf{\hat{H}}^{0}$ uses the LMFD algorithm. Since the PESs
of $\mathbf{\hat{H}}^{0}$ are decoupled and (with one exception shown
in Fig. \ref{fig:Adiab_basis}) only one PES is occupied initially,
the dynamics reduces to the simpler Born-Oppenheimer classical dynamics
of an ensemble of phase space points. In Subsec.\,\ref{sub:Two-different-coupled},
more general Hamiltonians and perturbations are studied.

\subsection{Nondiabaticity of quantum dynamics\label{sub:Diabatic-basis} }

\begin{figure*}
\includegraphics[width=\linewidth]{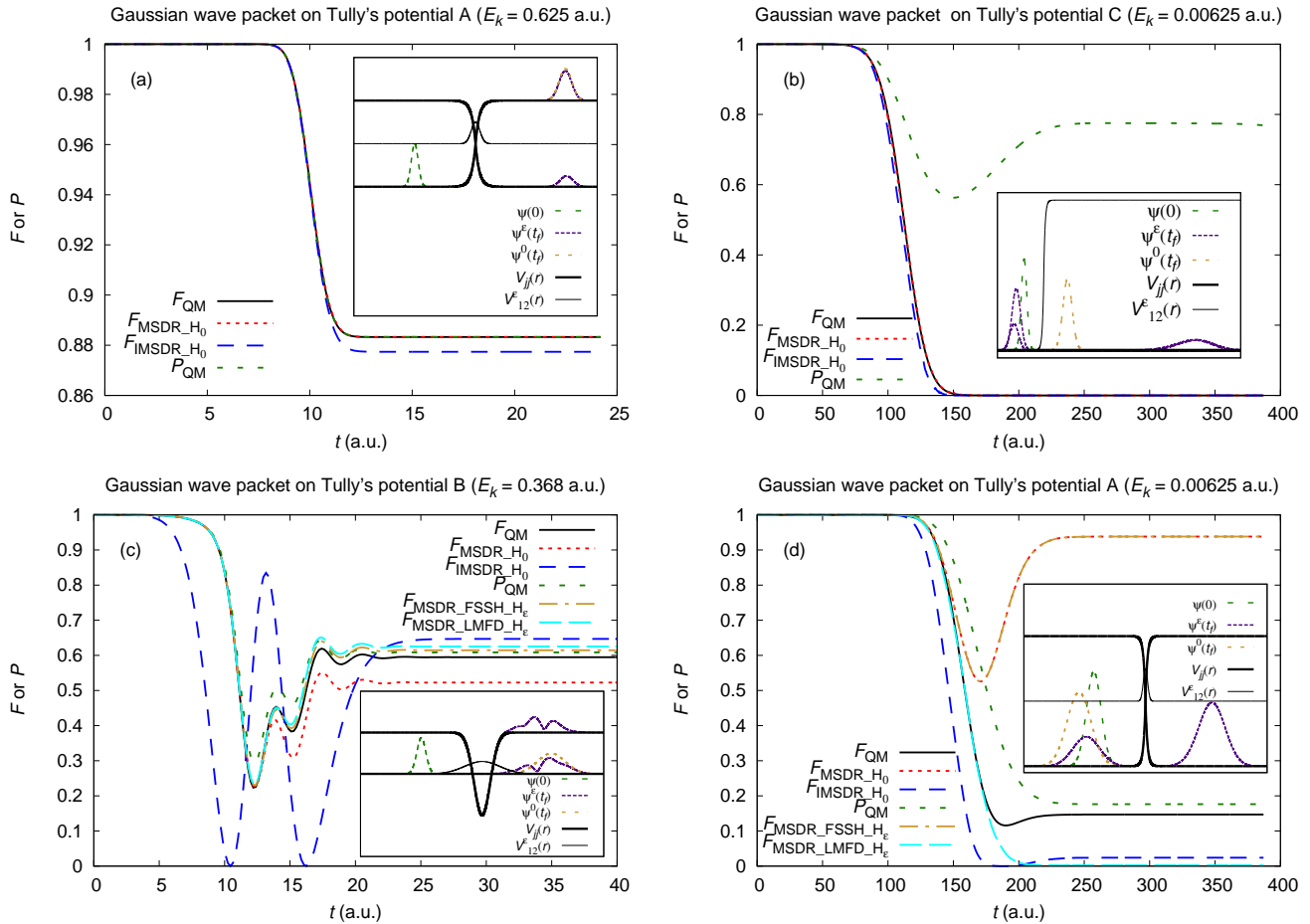}

\caption{Nondiabaticity of quantum dynamics. Panels\,(a)\protect\nobreakdash-(d)
compare the numerically exact quantum fidelity ($F_{\mathrm{QM}}$)
with extensions of the DR in the diabatic basis and with the quantum
survival probability ($P_{\mathrm{QM}}$). (a)\,Dynamics close to
the diabatic limit. (b)\,Dynamics very far from the diabatic limit.
(c)\,Dynamics in the intermediate range. (d)\,Dynamics in the region
where quantum tunneling is important. The insets show the diabatic
PESs $V_{jj}(r)$, diabatic coupling $V_{12}(r)$ as well as initial
{[}$\psi(0)${]} and final {[}$\psi^{0}(t_{f})$ and $\psi^{\epsilon}(t_{f})${]}
wave functions evolved with $\mathbf{\hat{H}}^{0}$ and $\mathbf{\hat{H}}^{\epsilon}$,
respectively. \label{fig:Diab_basis}}
\end{figure*}

In this Subsection, the fidelity criterion of nondiabaticity of the
quantum dynamics in the diabatic basis is studied together with the
IMSDR and MSDR approximations of fidelity. The IMSDR was already shown
to perform well only when the dynamics was close to the diabatic limit.\cite{Zimmermann2010}
On the other hand, as demonstrated here, MSDR performs well not only
in a broader range of problems but also for lower wave packet energies.
Comparisons of both methods with numerically exact quantum fidelity
($F_{\mathrm{QM}}$) in the diabatic basis can be found in Fig.\,\ref{fig:Diab_basis}.
Figure\,\ref{fig:Diab_basis}\,(a) shows results for the single
avoided crossing model of Tully.\cite{Tully1990} The initial wave
packet has high kinetic energy so that the dynamics is fairly close
to the diabatic limit. Thus both extensions of the DR work very well;
$F_{\text{MSDR}}$ is almost indistinguishable from $F_{\mathrm{QM}}$.
Very similar results were obtained for the other model potentials
when the wave packet had sufficient energy to cross the coupling region
almost diabatically. Figure\,\ref{fig:Diab_basis}\,(b) demonstrates,
on the extended coupling model of Tully,\cite{Tully1990} that the
survival probability $P_{\mathrm{QM}}$ itself is not always a good
indicator of nondiabaticity of the dynamics. Here, $F_{\mathrm{QM}}$
decays quickly to zero despite $P_{\mathrm{QM}}$ staying close to
unity. Indeed, the quantum dynamics on $\mathbf{\hat{H}}^{0}$ and
$\mathbf{\hat{H}}^{\epsilon}$ are very different. Both extensions
of the DR reproduce the decay of $F_{\mathrm{QM}}$ very accurately.
Using the double avoided crossing model of Tully,\cite{Tully1990}
Fig.\,\ref{fig:Diab_basis}\,(c) shows that MSDR can accurately
reproduce the fidelity behavior even far from the diabatic limit.
Not surprisingly, the IMSDR method fails here. For comparison, Fig.\,\ref{fig:Diab_basis}\,(c)
also shows two MSDR results obtained by exchanging the roles of $\mathbf{\hat{H}}^{0}$
and $\mathbf{\hat{H}}^{\epsilon}$. Since, in contrast to $\mathbf{\hat{H}}^{0}$,
$\mathbf{\hat{H}}^{\epsilon}$ is coupled, both the LMFD and FSSH
dynamics allow transitions between PESs: both are good approximations
of $F_{\mathrm{QM}}$. Finally, Fig.\,\ref{fig:Diab_basis}\,(d)
demonstrates that because the MSDR is a semiclassical method based
on classical trajectories, not permitting tunneling, the method has
to be applied with care. In the case shown, far from the diabatic
limit, the wave packet on $\mathbf{\hat{H}}^{0}$ reflects back from
the coupling region. In the MSDR this reflection results in ``rephasing''
of the trajectories leading to the rise of fidelity back to values
close to unity, in disagreement with the quantum result. Even if roles
of $\mathbf{\hat{H}}^{0}$ and $\mathbf{\hat{H}}^{\epsilon}$ are
exchanged, problems persist. In the quantum dynamics, a major part
of the wave packet on the coupled Hamiltonian $\mathbf{\hat{H}}^{\epsilon}$
passes through the coupling region with the aid of diabatic couplings.
When the LMFD is used with MSDR, this behavior and associated fidelity
decay are captured qualitatively. When the FSSH dynamics is used,
the dynamics is exactly identical to the dynamics on the uncoupled
Hamiltonian $\mathbf{\hat{H}}^{0}$ because all hops in the FSSH algorithm
are frustrated. This points out the importance of tunneling in this
relatively narrow region of wave packet energies. When the initial
kinetic energy is smaller than that shown in Fig.\,\ref{fig:Diab_basis}\,(d),
most of the wave packet bounces back even during quantum dynamics
on $\mathbf{\hat{H}}^{\epsilon}$ and the MSDR fidelity approaches
the quantum result. When the energy is somewhat higher, more trajectories pass
the barrier (or fewer hops in the FSSH algorithm on $\mathbf{\hat{H}}^{\epsilon}$
are frustrated) and as a consequence the MSDR approaches again the
quantum result. Note that the relatively good result of the IMSDR
is accidental since the method is not expected to work well so far
from the diabatic limit.

\subsection{Nonadiabaticity of quantum dynamics\label{sub:Adiabatic-basis} }

\begin{figure*}
\includegraphics[width=\linewidth]{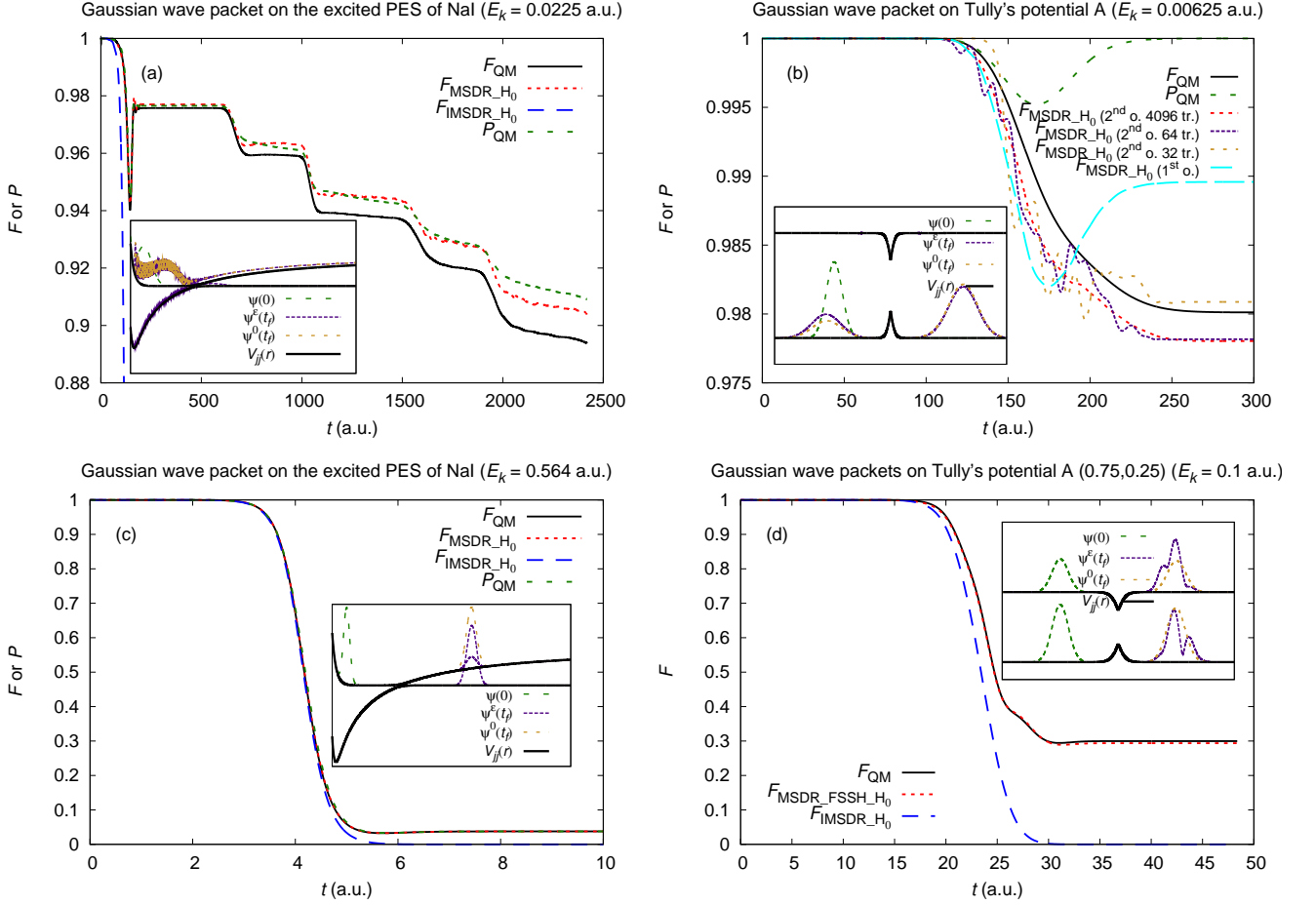}

\caption{Nonadiabaticity of quantum dynamics. Panels\,(a)\protect\nobreakdash-(d)
compare the numerically exact quantum fidelity ($F_{\mathrm{QM}}$)
with the extensions of the DR in the adiabatic basis and (when applicable)
with the quantum survival probability ($P_{\mathrm{QM}}$). (a)\,Dynamics
close to the adiabatic limit. (b)\,Convergence of the MSDR and importance
of the second order dynamics close to the adiabatic limit. (c)\,Dynamics
very far from the adiabatic limit. (d)\,A more general initial condition
with both PESs initially occupied. The insets show the adiabatic PESs
$V_{jj}(r)$ as well as initial {[}$\psi(0)${]} and final {[}$\psi^{0}(t_{f})$
and $\psi^{\epsilon}(t_{f})${]} wave functions evolved with $\mathbf{\hat{H}}^{0}$
and $\mathbf{\hat{H}}^{\epsilon}$, respectively.\label{fig:Adiab_basis}}
\end{figure*}

In this Subsection, the fidelity criterion of nonadiabaticity of the
quantum dynamics in the adiabatic basis is studied together with the
IMSDR and MSDR approximations of fidelity. As can be seen in Fig.\,\ref{fig:Adiab_basis},
in the adiabatic basis the IMSDR does not work at all, whereas the
MSDR works as satisfactorily as in the diabatic basis. Figure\,\ref{fig:Adiab_basis}\,(a)
shows results for the photodissociation dynamics of NaI close to the
adiabatic limit. A low energy wave packet oscillates on the excited
adiabatic PES, crossing the coupling region twice per period, each
time losing a bit of the excited PES population due to the transition
to the ground PES. The MSDR reproduces the associated fidelity decay
very well, whereas the IMSDR fails completely. Dynamics even closer
to the adiabatic limit is shown in Fig.\,\ref{fig:Adiab_basis}\,(b)
using Tully's single avoided crossing model A. First of all, note
that $F_{\mathrm{QM}}$ is again substantially different from the
survival probability $P_{\mathrm{QM}}$. Second, the figure also shows
that in certain cases the MSDR based on the first order nonadiabatic
mixed quantum-classical equation {[}Eq. \eqref{eq:mixed_Liouville}{]}
does not accurately reproduce $F_{\mathrm{QM}}$. Interestingly, the
agreement can be improved with the second order dynamics, which employs
the complete partially Winger transformed nonadiabatic Hamiltonian
\begin{equation}
\mathbf{H}_{\text{W}}^{\mathrm{A,comp}}=\mathbf{H}_{\text{W}}^{\text{A}}-i\hbar\frac{P}{M}\mathbf{F}-\frac{\hbar^{2}}{2M}\mathbf{F}\cdot\mathbf{F}.\label{eq:H_adiab_Wigner_complete}
\end{equation}
Strictly speaking such an approach goes beyond the MSDR defined by
Eqs. \eqref{eq:f_DR_D} and \eqref{eq:mixed_Liouville}. Nevertheless,
in the special case of a decoupled unperturbed Hamiltonian $\mathbf{\hat{H}}^{0}$
with only one occupied PES, the second order equation of motion can
be easily used instead of Eq. \eqref{eq:mixed_Liouville}, because
the increased order of dynamics affects only phases. Trajectories,
which are propagated with the decoupled Hamiltonian $\mathbf{\hat{H}}^{0}$,
stay unaffected since the contribution from the second order Poisson
brackets vanishes. The second-order Hamiltonian \eqref{eq:H_adiab_Wigner_complete},
however, cannot be derived by generalizing the approach used to derive
Eq. \eqref{eq:Liouville_adiab_1_local}. In that case, the mixed quantum-classical
equation \eqref{eq:mixed_Liouville} was derived from the Liouville-Von
Neumann equation in the diabatic basis and then transformed into the
adiabatic basis using Eq. \eqref{eq:d_rho_adiab} to yield Eq. \eqref{eq:Liouville_adiab_1_local}.
Generalizing this approach to the second order, no correction is obtained
to the first order Hamiltonian acting on the electronic degree of
freedom in the Lagrangian frame. Instead, we have used a generalization
to the second order of the approach from Ref. \onlinecite{Horenko2002},
where the mixed quantum-classical equation was obtained directly by
Wigner transforming the Liouville-Von Neumann equation expressed in
the adiabatic basis. The exact reason for the discrepancy between
the two approaches is not yet clear to us and will be a subject of
further investigation. (Some discrepancy is actually present already
in the first order.) In other examples that we have studied, the second
order correction does not significantly influence the results. Finally,
Fig.\,\ref{fig:Adiab_basis}\,(b) also demonstrates the convergence
of the MSDR with the number of trajectories showing that as few as
32 trajectories suffice for an accurate approximation of $F_{\mathrm{QM}}$.
(The convergence is similar for both the first order and the second
order dynamics.) For the DR\cite{Vanicek2006,Mollica2011a} and IMSDR,\cite{Zimmermann2010}
an exact formula exists for the number of trajectories as a function
of the statistical error and of fidelity. For the MSDR, such an exact
formula has not been derived, but similarly to the DR\cite{Vanicek2006,Mollica2011a}
and IMSDR,\cite{Zimmermann2010} more trajectories have to be used
for lower fidelity. Figure\,\ref{fig:Adiab_basis}\,(c) shows that
the MSDR usually retains its accuracy even far from the adiabatic
limit. Nevertheless, similarly to the diabatic basis, special care
should be taken in such cases.

In all examples discussed above, only one PES was occupied initially.
Yet the MSDR works also with more general initial conditions, as shown
in Fig.\,\ref{fig:Adiab_basis}\,(d) using Tully's single avoided
crossing model. Here 75 \% of the initial density is located on the
lower PES, and the rest on the upper one. Note that when more than
one PES of $\mathbf{\hat{H}}^{0}$ is occupied, one must watch out
for the intrinsic deficiencies of the underlying dynamical methods.
If the LMFD is used for propagation (not shown), each trajectory propagates
on an average PES, given by the weighted average of all occupied PES,
even outside of coupling regions. When the FSSH based algorithm is
used {[}as shown in Fig.\,\ref{fig:Adiab_basis}\,(d){]}, other
problems may occur in a similar situation, i.e., when wave packet
dynamics on the lower and upper PESs differ substantially outside
of coupling regions.

\subsection{Importance of an additional interaction in the Hamiltonian and accuracy
of an approximate Hamiltonian\label{sub:Two-different-coupled}}

Now we consider two more general applications of the MSDR. In the
first application, the MSDR is used to evaluate the importance of
an additional interaction term in the Hamiltonian. In the second application,
the method is employed to evaluate the accuracy of quantum dynamics
on an approximate non(a)diabatic Hamiltonian. In both cases, $\mathbf{\hat{H}}^{0}$
and $\mathbf{\hat{H}}^{\epsilon}$ are coupled and may differ in both
diagonal and coupling terms. Because even $\mathbf{\hat{H}}^{0}$
is coupled, the trajectories used in the MSDR do not anymore correspond
to simpler Born-Oppenheimer trajectories even when only one PES is
occupied initially\texttt{.}

\subsubsection{Importance of an additional interaction in the Hamiltonian}

The importance of an additional interaction in the Hamiltonian is
tested in Fig.\,\ref{fig:General_Hamiltonians}\,(a). This case
is similar to previously studied problems because $\mathbf{\hat{H}}^{0}$
and $\mathbf{\hat{H}}^{\epsilon}$ differ again only by the presence
of a coupling element; the difference is that $\mathbf{\hat{H}}^{0}$
is now coupled. Both diabatic Hamiltonians contain three PESs, of
which the lower two are identical to the PESs of Tully's single avoided
crossing model. The third PES is flat with constant energy $E=0.15\,\mathrm{a.u.}$
In both Hamiltonians, the lower two PESs are coupled by the same coupling
term $V_{12}$ as in the original single avoided crossing model. Additionally,
in $\mathbf{\hat{H}}^{\epsilon}$ (but not in $\mathbf{\hat{H}}^{0}$),
the highest two PESs are coupled with
\begin{equation}
V_{23}=V_{32}^{*}=\left(1+i\right)\left[C\exp\left(-DQ^{2}\right)\right],\label{eq:3surf_coupling}
\end{equation}
where $C=0.005$ and $D=1.0$. A more general complex form was chosen
to emulate the presence of spin-orbit coupling terms, which may also
be complex valued. As can be seen from the figure, the MSDR reproduces
the exact decay of fidelity accurately. Note that fidelity decays
significantly even though only approximately 1 \% of the probability
density ends up on the highest PES. In other words, this is yet another
example, where the survival probability is a poor measure of the importance
of couplings between PESs.

\subsubsection{Accuracy of an approximate Hamiltonian}

The last application of the MSDR that we consider is the evaluation
of the accuracy of an approximate Hamiltonian. An additional difficulty
related to this application is that the electronic basis sets used
to represent the two Hamiltonians may be different. In the adiabatic
basis set, electronic basis functions are determined at each space
point by the Hamiltonian itself. The diabatic basis, on the other
hand, is usually chosen to diagonalize the Hamiltonian at some fixed
nuclear configuration. To avoid difficulties, one may express both
Hamiltonians in the same basis set, but this is not always feasible
(at least not easily). Such problem occur, e.g., when one compares
two Hamiltonians computed on the fly using different electronic structure
methods. Since Hamiltonians used in electronic structure methods are
usually approximate, the adiabatic basis functions are not necessarily
identical. Moreover, in some methods, such as DFT, electronic basis
functions do not even have to be determined. Therefore, as we are
interested in the dynamics of nuclei and in the extent of electronic
transitions rather than in approximations underlying electronic structure
methods, we solve this problem by replacing our definition of fidelity
$F_{\text{QM}}$ with an alternative definition $F'_{\mathrm{QM}}$,
in which the overlap matrix of electronic basis functions is always
assumed to be the identity matrix. 

\begin{figure*}
\includegraphics[width=\linewidth]{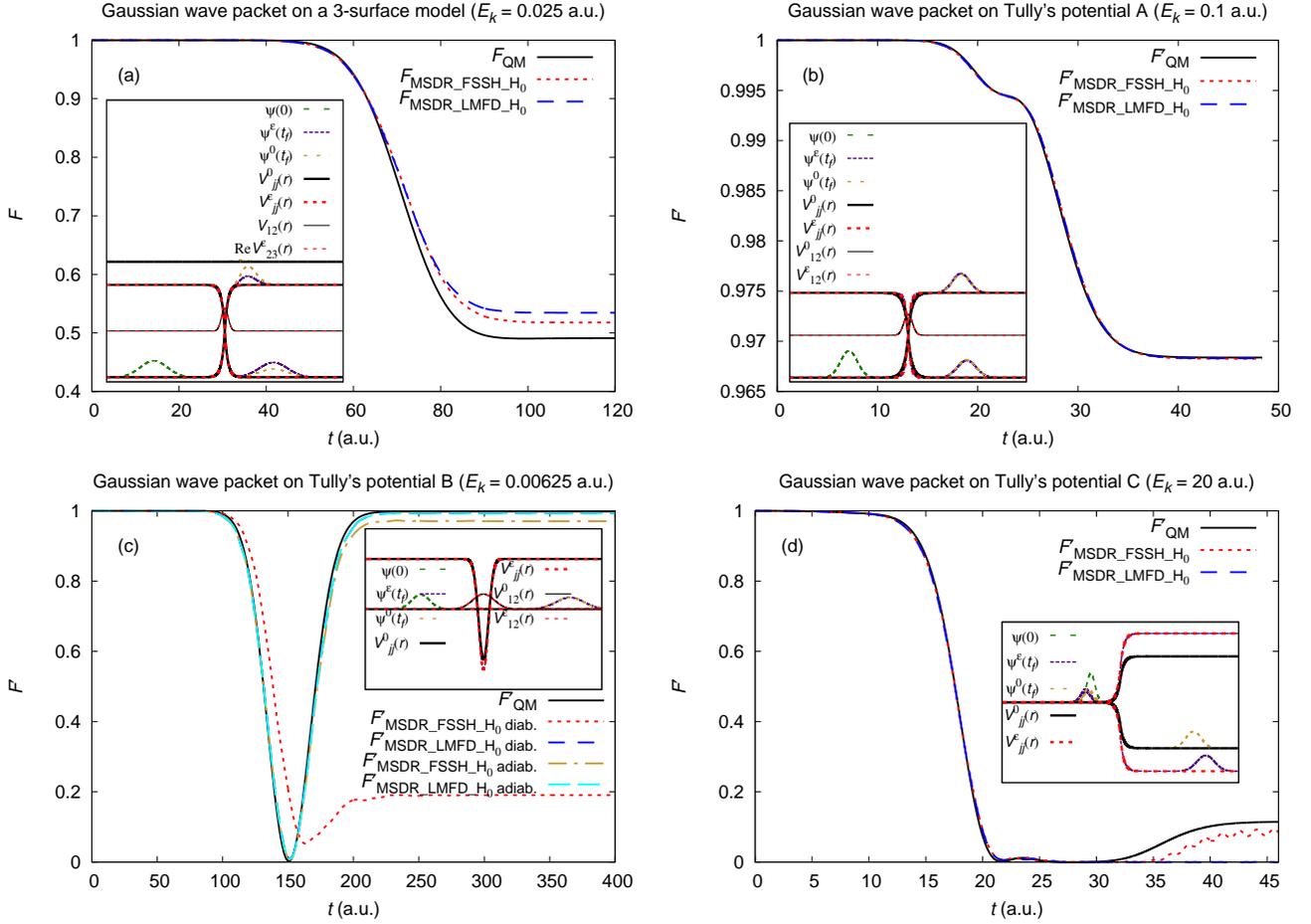}

\caption{(a)\,Fidelity as a measure of the effects of additional (e.g., spin-orbit)
couplings on the quantum dynamics. (b)-(d)\,Fidelity as a measure
of the accuracy of quantum dynamics on an approximate Hamiltonian.
Comparison of the numerically exact quantum fidelity ($F_{\mathrm{QM}}$
or $F'_{\mathrm{QM}}$ ) with extensions of the DR. The insets show
the diabatic or adiabatic PESs $V_{jj}(r)$, diabatic couplings $V_{12}(r)$
and $V_{23}(r)$ as well as the initial {[}$\psi(0)${]} and final
{[}$\psi^{0}(t_{f})$ and $\psi^{\epsilon}(t_{f})${]} wave functions
evolved with $\mathbf{\hat{H}}^{0}$ and $\mathbf{\hat{H}}^{\epsilon}$,
respectively.\label{fig:General_Hamiltonians}}
\end{figure*}

An ``approximate'' Hamiltonian $\mathbf{\hat{H}}^{\epsilon}$ was
created by perturbing one of the parameters in the analytical formula
for $\mathbf{\hat{H}}^{0}$ in the diabatic basis. (The perturbation
was transformed into the adiabatic basis when the dynamics was done
in the adiabatic basis). First, the effect of perturbing the slope
of PESs in the coupling region was studied using Tully's single avoided
crossing model in the diabatic basis {[}see Fig.\,\ref{fig:General_Hamiltonians}\,(b){]}.
The perturbation consisted in increasing the value of parameter $B$
in Eq.\,(21) in Ref.\,\onlinecite{Tully1990} by $50\%$. Although
electronic transitions are significant even on $\mathbf{\hat{H}}^{0}$
itself, the MSDR works very well with both the LMFD and FSSH dynamics.
Figure\,\ref{fig:General_Hamiltonians}\,(c) shows the decay of
fidelity due to an increased depth of the PES well in Tully's double
avoided crossing model. Calculations were performed both in the diabatic
and adiabatic basis. {[}Note that $F'_{\mathrm{QM}}$ may differ between
adiabatic and diabatic basis, but only in coupling regions. In the
case shown in Fig.\,\ref{fig:General_Hamiltonians}\,(c) the difference
is not significant.{]} To increase the depth of the well in $\mathbf{\hat{H}}^{\epsilon}$,
value of parameter $A$ in Eq.\,(23) in Ref.\,\onlinecite{Tully1990}
was increased by 10 \%. As can be seen in Fig.\,\ref{fig:General_Hamiltonians}\,(c),
the quantum result is reproduced very well by the MSDR based on the
LMFD in both basis sets. When the FSSH dynamics is used, the result
depends strongly on the basis. In the adiabatic basis, the MSDR reproduces
the quantum result quite well, whereas in the diabatic basis, the
method fails to follow $F'_{\mathrm{QM}}$ even qualitatively. Worse
performance of the FSSH dynamics in the diabatic basis is well known,\cite{Tully1998}
and in this specific case may be attributed mainly to the fact that
$40\%$ of hops in the diabatic basis are frustrated. 

Figure\,\ref{fig:General_Hamiltonians}\,(d) shows an opposite case,
where the MSDR works better with the FSSH than with LMFD. The model
used is Tully's extended coupling model in the adiabatic basis. To
introduce the perturbation, the coupling strength in $\mathbf{\hat{H}}^{\epsilon}$
(parameter $B$ in Eq.\,(24) in Ref.\,\onlinecite{Tully1990}) was
increased by $50\%$. The MSDR reproduces the initial decay of fidelity
irrespective of the dynamics used. By contrast, the subsequent rise
of fidelity caused by the partial reflection of the wave packet could
be reproduced only with the FSSH dynamics because no reflection occurs
in the mean field description.

\subsection{Computational details\label{sub:Computational-details}}

All quantum calculations in the diabatic basis set were performed
using the second order split-operator algorithm.\cite{Feit1983} Calculations
in the adiabatic basis were done either by transforming the quantum
state into the diabatic basis, propagating there, and transforming
back into the adiabatic basis, or directly with the second order Fourier
method.\cite{Kosloff1983} The LMFD or FSSH dynamics were done using
the second order symplectic Verlet integrator.\cite{Verlet1967}

\section{Discussion and Conclusions\label{sec:Discussion-and-Conclusions}}

Results presented above demonstrate that quantum fidelity is useful
as a quantitative measure of nondiabaticity or nonadiabaticity of
quantum dynamics. Moreover the MSDR, a semiclassical approximation
developed in this work, has proven to be a reliable yet very efficient
substitute for the expensive exact quantum dynamics calculation of
fidelity. The MSDR, similarly to the less accurate IMSDR,\cite{Zimmermann2010}
is a generalization of DR to non(a)diabatic dynamics. In addition
to quantum effects originating from the interaction of nuclei with
electrons, which are included in most mixed quantum-classical methods,
the MSDR includes also quantum effects originating from the interference
between mixed quantum-classical trajectories representing the evolution
of the initial density matrix.\texttt{ }Two approximate variants of
the MSDR were developed and studied numerically: the former uses the
LMFD, derived here, while the latter employs the FSSH dynamics. The
LMFD, although obtained in a different way, is equivalent to the nuclear
dynamics appearing in the nonadiabatic IVR,\cite{Sun1997,Ananth2007,Miller2009}
which is, in turn, nothing else than the Ehrenfest dynamics applied
separately to each classical phase space point representing the initial
density matrix. Both FSSH and the Ehrenfest method are relatively
simple and often used, thus both variants of MSDR can be easily implemented
into any FSSH or Ehrenfest dynamics code, especially for pure states. 

Several applications of the MSDR in nonadiabatic dynamics were presented.
First, the method was used to approximate a rigorous quantitative
measure of nondiabaticity or nonadiabaticity of the dynamics based
on quantum fidelity. As such the MSDR may be used to decide - before
running the quantum dynamics itself - which PESs or Hamiltonian terms
are important. Second, the method permits establishing the relative
importance of several interaction terms in a Hamiltonian. Third, generalizing
one of the applications of the original DR,\cite{Li2009,Zimmermann2010a}
the MSDR may be used to evaluate the accuracy of quantum dynamics
with an approximate non(a)diabatic Hamiltonian. Apart from these applications,
the MSDR could be used, in principle, to compute all quantities expressible
in terms of quantum fidelity or quantum fidelity amplitude, such as
various spectra. 

In Section\,\ref{sec:Results} we have demonstrated that for one
dimensional model systems the MSDR works often satisfactorily even
far from the diabatic or adiabatic limit. The method has yet to be
tested in multi-dimensional systems. Nevertheless, results obtained
with the original Born-Oppenheimer DR method demonstrate that the
convergence of the method is actually independent of the dimensionality
of a problem\cite{Mollica2011a} and that the accuracy does not deteriorate
with dimensionality.\cite{Li2009} Thus, we expect that the MSDR would
perform well especially in chaotic multi-dimensional systems such
as some molecules, provided that the underlying mixed classical-quantum
dynamics is a reasonable approximation to the quantum dynamics. Unfortunately,
this is not always the case. It is well known that the Ehrenfest dynamics
(and also the LMFD) can be qualitatively incorrect when coupling vanishes
after passing a coupling region and more PESs are occupied. In the
MSDR this problem is often less significant, because in many cases
after passing the coupling region fidelity does not decay anymore.
The FSSH dynamics also suffers from several problems besides the inaccuracies
caused by the classical description of nuclei, such as the problem
of ``excessive coherence'' related \texttt{(}similarly as for the
LMFD\texttt{)} to the fact that all matrix elements of the density
matrix attached to a trajectory evolve on the same PES. In many cases,
this can be alleviated by applying the ``decoherence'' correction.\cite{Fang1999,Zhu2004,Granucci2007,Granucci2010,Subotnik2011}
Including a similar correction into the MSDR method should be straightforward
and will be explored in future work.\texttt{ }Another possibility
how to overcome some deficiencies of the LMFD or FSSH dynamics would
be to use the MSDR together with one of the propagation methods attempting
to solve the mixed quantum-classical Liouville equation directly.\texttt{ }

\textit{Acknowledgement}

This research was supported by the Swiss NSF NCCR MUST (Molecular
Ultrafast Science \& Technology) and by the EPFL.

\end{document}